\documentclass[iop]{emulateapj}
\usepackage{graphicx,ifthen,url,float,hyperref,amsmath,textcomp}

\newcommand{\kms}{km\,s$^{-1}$}
\newcommand{\msun}{{\rm M$_\odot$}}
\newcommand{\rpm}{\raisebox{.2ex}{$\scriptstyle\pm$}}
\newcommand{\ang}{$\mbox{\AA}$}
\newcommand{\delv}{$\Delta V_{BA}$}

\shorttitle{MWC349A and B Are Not Gravitationally Bound}
\shortauthors{P.~M. Drew et al.}

\begin{document}


\title{MWC349A and B Are Not Gravitationally Bound: New Evidence}


\author{Drew, P.M.\altaffilmark{1}$^{,}$\altaffilmark{2}$^{,}$\altaffilmark{5}, Strelnitski, V.\altaffilmark{2}, Smith, H.A.\altaffilmark{3}, Mink, J.\altaffilmark{3}, Jorgenson, R.A.\altaffilmark{2}, and O’Meara, J.M.\altaffilmark{4}}

\altaffiltext{1}{University of Massachusetts at Amherst, 710 N. Pleasant St., Amherst, MA 01003}
\altaffiltext{2}{Maria Mitchell Observatory, 4 Vestal Street, Nantucket, MA 02554, USA}
\altaffiltext{3}{Harvard-Smithsonian Center for Astrophysics, 60 Garden Street, Cambridge, MA 02138, USA}
\altaffiltext{4}{Department of Chemistry and Physics, Saint Michael’s Col-
lege, One Winooski Park, Colchester, VT 05439}
\altaffiltext{5}{Department of Astronomy, The University of Texas at Austin, 2515 Speedway Blvd Stop C1400, Austin, TX 78712}



\begin{abstract}
The age and evolutionary status of MWC349A, the unique emission line star with maser and laser radiation in hydrogen recombination lines, remain unknown because the spectrum of the star is veiled by bright emission from the ionized disk and wind. The major argument for this massive ($>$10\,\msun) star being evolved is its association with a close-by (2.4 arcsec) companion, MWC349B, whose B0III spectrum implies an age of a few Myrs. However, newly obtained high-resolution spectra of MWC349B reveal a difference $\approx$ 35\,\kms\ in the radial velocities of the two stars, which makes their being gravitationally bound highly improbable. An estimate of the relative proper motion of the two stars seems to confirm this conclusion. This reopens the previously suggested possibility that MWC349A is a young massive star in a region of active star formation close to the Cyg OB2 association. MWC349B, which moves with a speed $\geq$ 35\,\kms\ relative to Cyg OB2, may be a runaway star from this association.
\end{abstract}


\section{Introduction}

MWC349 is a visual double star with a spectrally unclassified luminous emission-line component MWC349A (hereafter star A) and a B0III component MWC349B (star B), 2.4 arcsec apart \citep{Coh85}. The massive, ionized outflow from star A and its circumstellar disk is a source of bright radio emission with unique characteristics. It is the brightest known stellar source of cm radio continuum and the only known source of high-gain maser and infrared laser emission in hydrogen recombination lines [see a brief review in \citet{Str13}].

The mass of star A has been estimated by several authors to be from 10--15\,\msun\ \citep{Zha17} to 25--30\,\msun\ \citep{Pon94,Thu94}. The age and the evolutionary status of this massive star are uncertain. A direct spectral classification is impossible, because the absorption spectrum of the star's atmosphere is completely flooded by exceedingly strong emission lines from its ionized envelope. Some indirect arguments have been presented for both the star being a luminous massive pre-main sequence object and for it being an evolved supergiant [see \citet{Gva12} for a review of hypotheses].

One of the strongest arguments in favor of the latter hypothesis is the possible physical connection of star A with the less luminous and probably less massive star B, whose B0III spectrum indicates an age of $\sim$5\,Myrs. \citet{Coh85}, as well as \citet{Taf04} presented some evidence of a possible physical connection between the two stars, based on the shape of the radio continuum isophotes. However, \citet{Mey02} offer alternative explanations for the radio nebula shape and argue, based on their spectropolarimetry of the two stars and constraints from the interstellar polarization, that star B is appreciably farther from us than star A, and therefore they cannot be physical companions, although both stars are probably connected with the Cyg OB2 association. \citet{Man17} argue, based on their estimate of the reddening of the two stars, that star B cannot be much farther from us than star A and that it may be within the nebula of star A. \citet{Str13} drew attention to a possible physical connection of star A with a compact molecular cloud in the region of active star formation triggered by the mass outflow from Cyg OB2. If this connection is real, star A may be very young, and even may be the first case of a $>$10\,\msun\ Herbig AeBe star in a very short pre-main sequence phase, just after having dispersed most of the surrounding molecular cloud.

Using the high-resolution spectra of stars A and B obtained with the TRES spectrograph of the 1.5-m Tillinghast Reflector at the Fred L. Whipple Observatory, \citet{Dre16} came to the conclusion that the radial velocities of stars A and B are probably different by an amount several times greater than that allowed by a gravitationally bound system. However, the TRES spectrum of star B was heavily contaminated by the bright emission spectrum of star A, which caused considerable uncertainty in the measured radial velocities of its absorption lines. An independent confirmation came recently with the publication of the radial velocity of star B by \citet{Man17} and with our own measurement using a high quality spectrum obtained with the HIRES echelle spectrograph on the Keck I telescope.

In this paper we re-examine the results and conclusions briefly reported in \citet{Dre16}, present the confirming Keck/HIRES result and present a supporting argument based on an estimate of the relative proper motion of the two stars. The observations and reductions are described in \autoref{sec:obs} and the results -- in \autoref{sec:results}. In \autoref{sec:discussion}, we obtain the upper limit for the radial velocity difference between stars A and B if they are gravitationally bound and show that the observed velocity difference is several times greater than this limit, and thus that the two stars are very probably not gravitationally bound. We also obtain a preliminary estimate of the relative proper motion of the two stars, which seems to confirm their high relative velocity. Section \ref{sec:conc} summarizes our conclusions.

\section{Observations and Data Reduction}
\label{sec:obs}
\subsection{The Tillinghast/TRES data}
\label{sub:Tres}

The spectra of both stars were obtained on 2014 October 31 using the TRES fiber-fed echelle spectrograph on the 1.5-m Tillinghast Reflector at the Fred L. Whipple Observatory. The spectral resolving power was 44,000 (6.8\,\kms). The spatial resolution, limited by the input of the fiber optics feed with a diameter of 2.4 arcsec, was comparable to the angular separation of the stars. As a result, the spectrum of star B was strongly contaminated by very bright emission lines of star A.

We used the {\sc iraf}\footnote{{\sc iraf} is distributed by the National Optical Astronomy Observatories, which are operated by the Association of Universities for Research in Astronomy, Inc., under} package {\sc splot} to analyze the spectra. In order to extract the uncontaminated absorption spectrum of star B a simple de-contamination procedure was applied based on the assumption that the measured intensity $I_{\lambda B}$ at each wavelength $\lambda$ of star B's spectrum is the sum of the intensity due to this star, $I^{0}_{\lambda B}$, and a fraction $f$ of $I_{\lambda A}$ -- the observed intensity of the spectrum of star A.  Thus, the uncontaminated intensity is given by the equation: $I^{0}_{\lambda B} - f\cdot I_{\lambda A}$. The unknown ``contaminating fraction'' $f$ was determined visually for the wavelength interval comprising each of the measured lines by gradually increasing the value of $f$ and looking for the appearance of the absorption profile that would show neither signatures of contaminating emission, nor signatures of ``over-subtraction'' of the emission (i.e.\:the reversed pattern of the contaminating emission line). Since most of star A's emission lines have a characteristic double-peaked profile, it was relatively easy to recognize the signatures of both residual contamination and ``over-subtraction.''  Still, this procedure was plagued by considerable uncertainty in the centroid radial velocity of each measured absorption line.

\subsection{The Keck/HIRES data}
\label{sub:Keck}
The Keck/HIRES echelle spectrograph \citep{Vog94} data were acquired on 2017 May 19 and 20. With the seeing of $\approx$0.8\,arcsec, we were able to obtain uncontaminated individual spectra of stars A and B by placing the slit perpendicular to the line connecting the images of the stars. A total exposure time of 600\,s on MWC349B using the C1 decker resulted in a signal-to-noise ratio $\rm S/N \approx 30$ per 1.3\,\kms\,pixel, with an instrumental resolution of $\rm FWHM = 6.25$\,\kms. The HIRES data were reduced using standard {\sc XIDL}\footnote{http://www.ucolick.org/$\sim$xavier/IDL/} reduction packages.

\section{Results}
\label{sec:results}
\begin{figure}
    \centering
    \includegraphics[width=0.99\columnwidth]{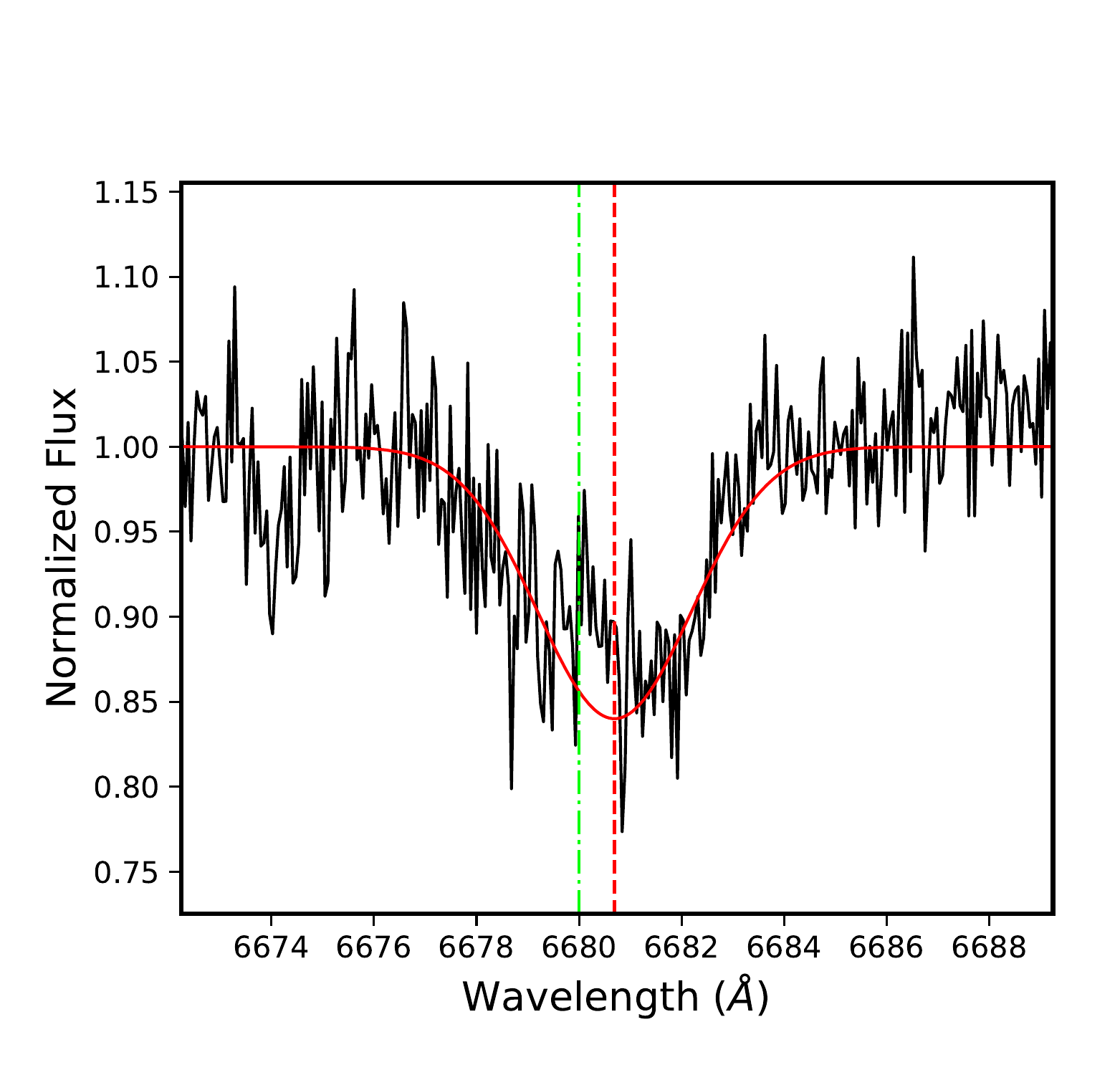}
    \caption{Decontaminated Tillinghast/TRES spectrum of star B over the region containing HeI line with vacuum rest wavelength 6680.00\,\ang. This wavelength is shown by green dot-dashed line. The best-fit Gaussian is shown by the red solid line. The red dashed line marks the centroid of the Gaussian, at $6680.69\rpm0.02$\,\ang, which corresponds to the heliocentric radial velocity of $+31\rpm1$\,\kms}
    \label{fig:fig1}
\end{figure}

Using the decontamination procedure described in \autoref{sub:Tres}, we measured the heliocentric radial velocities of three HeI absorption lines in the Tillinghast/TRES spectrum of star B. The adopted vacuum rest wavelengths of the three lines (5877.27\,\ang, 6680.00\,\ang, and 7067.17\,\ang) are the weighted averages of the strongest components of each line taken from the NIST database \citep{Kra15}. Figure \ref{fig:fig1} shows, as an example, the decontaminated profile of the 6680.00\,\ang\ line, whose centroid is at $6680.69\rpm0.02$\,\ang, corresponding to the heliocentric radial velocity V$_{B}($6680$) = +31\rpm1$\,\kms. The mean value of the heliocentric radial velocity over the three measured lines is given in the third column of Table 1. The error of the mean value (estimated from the scatter of the individual values of V$_{B}$) is considerably larger than the formal errors of the centroid of the fitting Gaussian for each line because of the uncertainty in determining the optimal value of the contamination fraction $f$ in the decontamination procedure.

\begin{figure}
    \centering
    \includegraphics[width=0.99\columnwidth]{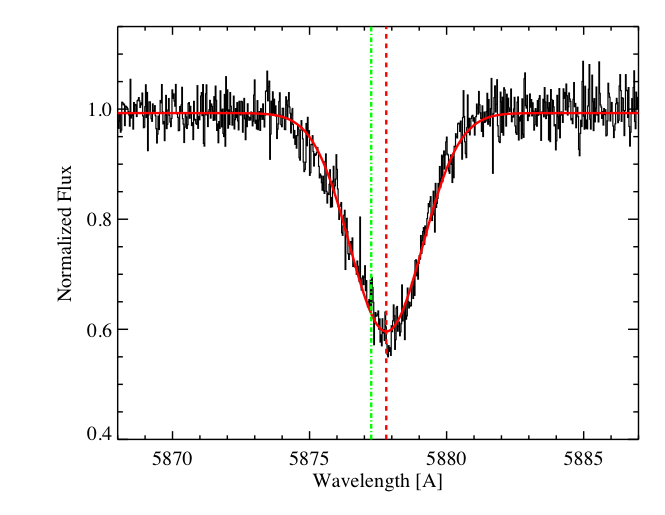}
    \caption{Keck/HIRES spectrum of star B over the region containing the HeI 5877.27\,\ang\ line. The rest wavelength of the line is indicated by the green dot-dashed line. The best-fit Gaussian is shown by the red solid line, with the red dashed line marking its centroid (see text and \autoref{tab:table1}).}
    \label{fig:fig2}
\end{figure}

We determined the radial velocity of star B with the Keck/HIRES spectrum using the 5877.27\,\ang\ HeI absorption line\footnote{The HIRES spectra of both stars will be discussed in further detail in an upcoming paper (Jorgenson et al., in preparation).}. After continuum fitting and normalization of the spectrum, a least-squares Gaussian fit to the absorption line gave a best-fit centroid of $5877.798\rpm0.014$\,\ang, or V$_{B}($HEL$) = +27.2\rpm1.5$\,\kms\ (\autoref{fig:fig2} and \autoref{tab:table1}).

There is no direct information on the radial velocity of star A because of the lack of its atmospheric absorption spectrum. However, its probable radial velocity can be estimated from the optical and radio data on the outflow and the disk of the star \citep[e.g.][]{Are16,Gor01}. Here we adopt star A's systemic velocity based on the SMA interferometry of the circumstellar disk in masering H$30\alpha$ and H$26\alpha$ lines \citep{Zha17}: V$_{A}($LSR$) = + 7.2\rpm0.2$\,\kms\ (the error estimate is based on the difference of $\approx$0.2\,\kms\ between the velocity values obtained with the two lines). For the ``standard'' solar motion adopted at SMA, the heliocentric velocity corresponding to this LSR velocity is V$_{A}($HEL$) = -9.9\rpm1.0$\,\kms\ (N. Patel, private communication; we increased the probable error to account for the small uncertainty of the LSR to HEL transformation). This value of V$_{A}$(HEL) is in agreement with the heliocentric radial velocity of the ionized envelope of the star as measured by \citet{Are16} using the forbidden optical lines of [OI] and [CaII]: V$_{\rm env}($HEL$) = -9\rpm2$\,\kms.

Using the above values of V$_{B}$(HEL) obtained with the two spectrographs, we calculated the values of {\delv} = V$_{B}$(HEL)$\,-\,$V$_{A}$(HEL). They are shown in the fourth column of \autoref{tab:table1}, together with their estimated errors.

\begin{table*}[]
    \centering
    \begin{tabular}{|c||p{1.8cm}p{1.8cm}p{1.8cm}p{1.8cm}p{3.1cm}|}
    \hline
         {\sc Instrument} & {\sc Spatial Res., arcsec} & {\sc Spectral Res., {\kms}} & {\sc $V_{B}$(HEL), {\kms}} & {\sc{\delv}, {\kms}} & {\sc Comments}\\
         \hline\hline

         Tillenghast/TRES & 2.4 & 6.8 & 30$\rpm$10 & 40$\rpm$10 & Average over three lines\\
         \hline
         
         Keck/HIRES & 
         0.7 &
         6.25 & 
         27.2$\rpm$1.5 &
         37$\rpm$2&
         A single line\\
         \hline
         
         CFHT (Manset et al. 2017) &
         - &
         5 &
         26$\rpm$1 &
         36$\rpm$1 &
         Average over two lines\\
         \hline
         
    \end{tabular}
    \caption{Comparison of three sources of {\delv}}
    \label{tab:table1}
\end{table*}

\section{Discussion}
\label{sec:discussion}
Given the much larger uncertainty of the Tillinghast/TRES result, the agreement between the values of {\delv} obtained with this spectrograph and with Keck/HIRES is remarkably good. These results are also corroborated by the recently published radial velocity of star B based on the measurement of two HeI lines in a spectrum obtained with the 3.6-m Canada-France-Hawaii Telescope \citep{Man17}, see \autoref{tab:table1}. Thus, we can firmly conclude from three independent results that the radial velocity difference of stars B and A is not less than $\approx$35\,\kms.

We now show that this velocity difference is much larger than the difference allowed for a gravitationally bound pair of stars with the probable parameters of MWC349A and B and the probable distance to them.

The radial velocity of a component of a binary (let it be component 1) is given by the standard equation:
\begin{gather}
V_{1} = V_{0} + K_{1}[\cos(\omega + \nu) + e\,\cos(\omega)],
\end{gather}
where V$_{0}$ is the radial velocity of the binary's center of mass, $\omega$ is the argument of periapsis of the orbit, $\nu$ is the true anomaly of the star within the orbit, $e$ is the orbit's eccentricity,
\begin{gather}
K_{1} = \frac{\sqrt{G(m_{1}+m_{2})}\,a_{1}\sin(i)}{\sqrt{a^{3}(1-e^{2})}},
\end{gather}
is the radial velocity semi-amplitude, $i$ is the inclination of the orbital rotation axis to the line of sight, $a \equiv a_{1}+a_{2}$ is the sum of the semi-major axes of the orbits of components 1 and 2, $m_{1}$ and $m_{2}$ are the masses of the components, and $G$ is the gravitational constant.

With similar expressions for $V_{2}$ and $K_{2}$ for component 2 and with the usual radial velocity sign convention (positive for recession and negative for approach), the observed absolute value of the difference of radial velocities of the components,
\begin{gather}
\Delta V_{12} \equiv | V_{1} - V_{2} |,
\end{gather}
is maximum when: (1) the plane of the orbits is seen edge-on ($i=90^{\circ}$); (2) the stars reach their highest orbital speeds, i.e.\:each of them is at the periapsis of its orbit ($\nu=0$); (3) the velocity vectors are parallel to the line of sight, i.e.\:the line of the major axes of the orbits is perpendicular to the line of sight ($\omega=0$). For this configuration, one gets from Equations (1)--(3), as well as two equations for component 2 analogous to Equations (1) and (2):
\begin{gather}
\Delta V_{12} = (1+e) \sqrt{\frac{G(m_{1}+m_{2})}{a(1-e^{2})}}.
\end{gather}

The linear distance between the stars when they are at their periapsides is $\delta = a(1-e)$. Taking into account: (1) that in the orbit configuration described above the line connecting the components lies in the plane of the sky, and thus $\delta = \theta \cdot D$ with $\theta$ being the observed angular separation of the components and D being the distance to the object, and (2) that $\theta \cdot D < \delta$ for any deviation of the line connecting the components from the plane of the sky, we finally get the upper limit:
\begin{gather}
\Delta V_{12} = \sqrt{\frac{G(1+e)(m_{1}+m_{2})}{\theta \cdot D}}\leq\sqrt{\frac{2G(m_{1}+m_{2})}{\theta \cdot D}},
\end{gather}
where the second inequality corresponds to extremely elongated orbits ($e\to1$). Substituting the current estimates of the masses of the components (quite probably the upper limits in both cases), $m_{1}\approx30$\,\msun\ and $m_{2}\approx20$\,\msun\ and also $\theta \approx2.5$\,arcsec (see below about the possible increase of the visible separation from 2.4 to 2.8\,arcsec in the last several decades), $D\approx1.5$\,kpc (averaging the extreme estimates of 1.7\,kpc \citep{Mas91} and 1.2\,kpc \citep{Coh85}), we get $\Delta V_{12} \leq 4.9$\,\kms. Note that this is a strict upper limit because we assumed a very special orientation of the orbits relative to the observer and a very special time (periapsis) for the stars in their motion within extremely elongated elliptical orbits. If, for example, the orbits were almost circular ($e \to 0$), the same special orientation would result in $\Delta V_{12} \leq 3.4$\,\kms.

Thus, the observed difference of radial velocities of stars A and B is much larger than the expected upper limit of the difference in radial velocity for two gravitationally bound stars. The possibility that the large radial velocity difference is due to the orbital motion (if star B is itself a binary) seems quite improbable. The period $P$ of a binary can be presented in the form:
\begin{equation}
\begin{gathered}
\frac{P}{\rm days} = 0.97 \times 10^{7} \left(\frac{m_{2}^{3}/(m_{1}+m_{2})^{2}}{M_{\odot}}\right)\left(\frac{K_{1}}{\rm km \,s^{-1}}\right)^{-3} \\ \times\left(\frac{\sin^{3}i}{(1-e^{2})^{1.5}}\right)
\end{gathered}
\end{equation}
Substituting the observed {\delv} $\approx$ 35\,\kms\ for $K_{1}$, and supposing that the mass of the component 2 in the star B binary is comparable or smaller than the probable mass of component 1 (star B, $\sim$20\,\msun), and that the eccentricity of the orbit is not too large, it is found from equation (7) that the probable period of the binary is $\sim$1000 days or less. During an orbital period, the radial velocity of star B (and thus {\delv}) must change by at least $\approx 35 \cdot 2 \approx 70$\,\kms. Yet, the three independent spectra taken with the intervals of 413 and 929\,days showed practically the same values of {\delv}, which for the earliest and the latest spectra, separated by 1342\,days, coincided within $\approx$6\% uncertainties of the obtained {\delv}. This corresponds to only $\approx$3\% of the expected radial velocity amplitude, were star B a binary. Such a coincidence is highly improbable, all the more so when one considers a very close {\delv} value obtained (although with a larger uncertainty) with the TRES spectrograph at a random moment between the earliest and the latest observation.

Thus, with a high probability, the large radial velocity difference of stars A and B is due to the permanent velocity difference of their centers of mass, and since it surpasses by much the difference allowed for two gravitationally bound stars, we conclude that stars A and B are not gravitationally bound.

In principle, this fact does not exclude that the stars were bound in the past but then disintegrated, like in the scenario proposed by \citet{Gva12}, in which the originally triple star system dissolves into a close binary (star A and its hypothetical close companion) and an escaping star B. Garamadze and Menten hypothesized that the initial triple system was ejected from the core of the Cyg OB2 association about 5\,Myrs ago after which, relatively recently, the disintegration of the system took place. However, two facts are against the first step of this scenario (escape from the core of Cyg OB2): (1) the closeness of the heliocentric radial velocity of MWC349A ($-9\rpm1.0$\,\kms; see above) to the systemic radial velocity of Cyg OB2 ($-10.3\rpm0.3$\,\kms; \citet{Kim07}), and (2) the apparent association of MWC349A with a compact molecular cloud having the same radial velocity \citep{Str13}. In contrast to star A, the radial velocity of star B differs from the radial velocity of the association by $\approx$35\kms, which makes it a good candidate runaway star from this association. 

A decisive check of our conclusion that the two stars are not gravitationally bound may be provided by a measurement of their proper motion. If the difference of tangential velocities is equal to or greater than the observed difference of radial velocity ($\sim$30\,\kms), the relative proper motion of the two stars, at their distance $\sim$1.5\,kpc, should be equal or greater than $\sim$4\,milliarcsec per year.

So far, the best estimate of the proper motion of star A has been done by \citet{Rod07} radioastronomically: $\mu_{\alpha}\cos\delta = -3.1 \rpm 0.5$\,mas\,yr$^{-1}$; $\mu_{\delta} = -5.3 \rpm 0.5$\,mas\,yr$^{-1}$.  These values are in agreement with the typical proper motion parameters for the Cyg X star forming complex \citep[e.g.][]{Ryg12}, although the uncertainty of measurements still allow for a peculiar motion of star A with a speed of several tens of \kms. A preliminary estimate of the proper motion of star B relative to star A can be done by comparing the separation of the two stars ($d_{1} = 2\arcsec.4 \rpm 0\arcsec.1$ in position angle $p_{1} = 280^{\circ} \rpm 2^{\circ}$ obtained in 1983 July \citep{Coh85} with the positions of the stars determined in the CDSS survey in 2003 September: RA$_{A} = 308^{\circ}.189657$; DEC$_{A} = 40^{\circ}.660171$; RA$_{B} = 308^{\circ}.188674$; DEC$_{B} = 40^{\circ}.660388$, with an uncertainty of $\rpm0\arcsec.05$ in both coordinates \citep{Ahn12}. This gives the separation $d_{2} = 2\arcsec.79\rpm0\arcsec.05$ in the position angle $p_{2} = 286^{\circ}\rpm1^{\circ}$. It is easy to see then that the observed angle covered by star B in the plane of the sky, in the reference frame of star A, is $0\arcsec.5\rpm0\arcsec.2$. This was covered in 20\,years, thus the relative proper motion was $25\rpm10$\,mas\,yr$^{-1}$ which, at a distance of $\approx$1.5\,kpc, corresponds to a relative transverse velocity of the two stars approximately $200\rpm100$\,\kms. This large relative transverse velocity seems to corroborate our conclusion that stars A and B are not gravitationally bound. However, this crude estimate is not based on a consistent astrometric study, which is still needed.

\section{Conclusion}
\label{sec:conc}
Recent measurements of absorption lines in the spectrum of MWC349B, the visible companion of the peculiar emission-line star MWC349A, reveal a large difference in radial velocities between the two stars. The difference is much larger than the theoretical upper limit for a gravitationally bound binary system. This makes it improbable that the two stars are physically connected, which reopens the previously suggested possibility that the $>$10\,\msun\ star A is very young and may even still be in its short pre-main sequence phase. With its high, positive radial velocity, star B may be a runaway star from the Cyg OB2 association. A preliminary estimate of the relative proper motion of the two stars seems to support these conclusions, but a targeted astrometric study is still needed to provide a decisive measurement and confirmation.

\section{acknowledgements}
We gratefully acknowledge David Latham for his help in organizing the TRES observations and for important suggestions on the draft of the paper. P.D. acknowledges financial support by the NSF REU grant 0851892, by the Nantucket Maria Mitchell Association, and by the University of Texas at Austin College of Natural Sciences. Some of the data presented herein were obtained at the W. M. Keck Observatory, which is operated as a scientific partnership among the California Institute of Technology, the University of California and the National Aeronautics and Space Administration. The Observatory was made possible by the generous financial support of the W. M. Keck Foundation. The authors wish to recognize and acknowledge the very significant cultural role and reverence that the summit of Mauna Kea has always had within the indigenous Hawaiian community. We are most fortunate to have the opportunity to conduct observations from this mountain. Finally, we thank the anonymous referee for critical remarks that led to a considerable improvement of the text.

~

\clearpage

\end{document}